\documentclass{article}
\usepackage[utf8]{inputenc}
\usepackage{booktabs}
\usepackage{graphicx}
\usepackage{geometry}
\usepackage{amsfonts}
\usepackage{physics}
\usepackage{float}
\usepackage{dynkin-diagrams}
\usepackage{amsmath,amssymb,amsthm,amscd,mathrsfs,cite}
\usepackage[hidelinks]{hyperref}

\usepackage{subcaption}
\usepackage[export]{adjustbox}    
\usepackage{xcolor}
\usepackage{listings}

\lstdefinestyle{mstyle}{
  basicstyle=\ttfamily\small,
  breaklines=true,
  breakatwhitespace=false,
  columns=fullflexible,
  keepspaces=true,
  showstringspaces=false,
  frame=single,
  framerule=0.4pt,
  numbers=left,
  numberstyle=\tiny,
  numbersep=8pt,
  tabsize=2,
  commentstyle=\itshape\color{gray},
  keywordstyle=\bfseries,
  stringstyle=\color{teal}
}

\usepackage{slashed}                      
\usepackage{tikz}                          
\usepackage{url}

\def \be {\begin{equation}}
\def \ee {\end{equation}}
\def \ba {\begin{aligned}}
\def \ea {\end{aligned}}
\def \bea {\begin{eqnarray}}
\def \eea {\end{eqnarray}}
\newcommand{\re}{{\mathrm{e}}}
\newcommand{\ri}{{\mathrm{i}}}
\newcommand{\rd}{{\mathrm{d}}}
\newcommand{\rH}{{\mathrm{H}}}

\begin{document}

\begin{titlepage}
\begin{flushright}
USTC-ICTS/PCFT-26-50\\
TIT/HEP-712
\end{flushright}
\vspace{0.5cm}
\begin{center}
{\Large \bf 
Partition Functions of Hermitian and PT-Symmetric Oscillators from Integrable Models}
\lineskip .75em
\vskip 2.5cm
{ Hongfei Shu$^{a,}$\footnote{ shu@zzu.edu.cn, shuphy124@gmail.com
} and Jingjing Yang$^{b,c,d,}$\footnote{jj150632@mail.ustc.edu.cn }
}
\vskip 2.5em
 {\normalsize\it 
$^{a}$Institute for Astrophysics,
Zhengzhou University, Zhengzhou, Henan 450001, China\\
$^{b}$Interdisciplinary Center for Theoretical Study,
University of Science and Technology of China, Hefei, Anhui 230026, China\\
$^{c}$Peng Huanwu Center for Fundamental Theory, Hefei, Anhui 230026, China\\
$^{d}$Department of Physics, Institute of Science Tokyo, Tokyo, 152-8551, Japan
}
\vskip 3.0em
\end{center}
\begin{abstract}

   We develop an ODE/IM-based formulation for the thermal partition function and the spectral zeta function of the homogeneous Hermitian and PT-symmetric oscillators. For both classes of systems, the quantization condition can be expressed using the counting function $a(E)$, which can be solved via the Destri--de Vega equation of the integrable model. We then express the partition function and spectral zeta function as contour integrals involving the counting function, thereby providing a direct bridge between quantum spectral functions and integrable models.
  \end{abstract}
\end{titlepage}

\newpage

\section{Introduction}

    The partition function is one of the most fundamental quantities in quantum mechanics and quantum field theory. The partition function simultaneously probes the low-lying spectrum in the low-temperature regime and the asymptotic distribution of highly excited states in the high-temperature regime, which provides a good playground to study the perturbative and non-perturbative aspects of a quantum system. However, even when the spectral problem reduces to a one-dimensional Schr\"odinger equation, obtaining the exact partition function remains a highly nontrivial problem. In the Euclidean formulation, the partition function $Z(\beta)$ is represented by a path integral over periodic trajectories in Euclidean time \cite{FeynmanHibbs1965}. The semi-classical expansion is organized around periodic solutions of the equation of motion, such as the instantons and bounces \cite{Coleman1977,Coleman1985,Zinn-Justin:2004vcw}. The Lefschetz thimble decomposition organizes the trans-series and the cancellation of ambiguities \cite{Witten:2010zr}.

    Recently, a non-trivial relation between the saddle point description and the exact WKB quantization condition has been found \cite{Sueishi:2020rug,Sueishi:2021xti}. In the exact WKB approach, one can derive the exact quantization condition $D(E)=0$ of the corresponding QM with a given boundary condition \cite{Voros1983,Silverstone85,DDP93,DDP97,DP99,Sueishi:2020rug}, where $D(E)$ is denoted as the spectral determinant.  The partition function is obtained via the contour integral encircling the spectrum $Z(\beta)=\oint \re^{-\beta E}\frac{\partial}{\partial E}\log D(E)\rd E$. 
    This form has been rederived using the quantum version of the Hamilton-Jacobi formalism of the phase space path integral \cite{Ture:2024nbi}. This procedure has unified two well-known non-perturbative approaches of QM, the path integral approach and the exact WKB approach. However, the determinant $D(E)$ is usually written in terms of the Borel resummed WKB periods. This makes the quantization condition exact, but requires order-by-order computation, resummation, and analytic continuation of the all-order WKB periods.

    On the other hand, a correspondence, known as the ODE/IM correspondence, between one-dimensional QM and the quantum integrable model has been found \cite{Dorey:1998pt,Bazhanov:1998wj}, where the exact WKB periods are found to satisfy the TBA equations \cite{Ito:2018eon,Ito:2025pfo}. The TBA equations provide a non-trivial method to compute the Borel resummed WKB periods without computing the WKB expansion order by order. Combining the TBA equations and the exact quantization condition provides a powerful method to solve the spectral problem \cite{Ito:2018eon, Emery:2020qqu, Ito:2023cyz}. For the homogeneous potentials, a Destri--de Vega (DdV)-type nonlinear integral equation \cite{Destri:1992ey,Bazhanov:1996dr} gives a particularly economical description of the radial spectral problem by using the counting function $a(E)$ \cite{Dorey:1999uk,Dorey:2007zx}.
    
    In this paper, we apply this framework to the homogeneous Hermitian and PT-symmetric oscillators, $V(x)=x^{2M}, -(\ri x)^{2M}$. More general potentials will be left for future work. Choosing relevant boundary conditions, the quantization condition can be expressed in terms of the counting function by $1+a(\pm E)=0$, which provides a contour integral representation for the partition function. Our construction provides an IM description for the partition function and zeta functions, and avoids an order-by-order calculation of the spectrum. We evaluate these quantities for representative Hermitian and PT-symmetric homogeneous oscillators and compare the results with direct spectral summation and known exact-WKB results. Furthermore, we find that the high-temperature expansion of the partition function is related to the integrals of motion arising from the large $E$ expansions of the counting function $a(E)$. This gives the exact expressions for the residues and special values of the spectral zeta function.

This paper is organized as follows. In section \ref{sec:review}, we review some basics of the spectral functions and the ODE/IM correspondence. The partition function $Z(\beta)$, zeta function $\zeta_{\rH}(s)$, and the counting function $a(E)$ will be introduced. In section \ref{sec:herm}, we compute the spectral functions for the Hermitian homogeneous anharmonic oscillator based on ODE/IM correspondence and compare our results with direct spectral summation and the known exact-WKB result. The relation between high-temperature expansion of the partition function and the integrals of motion arising from the large $E$ expansions of the counting function will also be shown.
In section \ref{sec:PT}, we study the PT-symmetric homogeneous oscillator. In the last section, we conclude and discuss the possible future directions.

\section{Spectral functions and DdV equation}
\label{sec:review}
Let us consider the one-dimensional Schr\"odinger equation
\begin{equation}
    \left(\rH-E\right)\psi(x)=0,
\end{equation}
where $\rH=-\frac{\rd^2}{\rd x^2}+V(x)$. We suppose the potential is bounded from below and confined at infinity. The Hamiltonian admits discrete, positive, and real eigenvalues: $E_0<E_1<\cdots<E_n<\cdots$. The finite-temperature partition function is defined by
\begin{equation}
    Z(\beta)=\operatorname{Tr} e^{-\beta \rH}=\sum_{n=0}^{\infty} e^{-\beta E_n} \quad(\operatorname{Re} \beta>0),
\end{equation}
where $\beta$ denotes the inverse temperature. The partition function packages the entire spectrum into an analytic function. At low temperature/large $\beta$, the partition function is dominated by the ground state energy $E_0$. We thus could extract $E_0$ from the partition function by
\begin{equation}\label{eq:E0-par}
    E_0=-\lim_{\beta\to+\infty}\frac{1}{\beta}\log Z(\beta).
\end{equation}
At high temperature/small $\beta$, the excited states will also contribute, whose asymptotics is governed by the phase-space volume \cite{Wigner:1932eb, Kirkwood:1933cn, Jizba:2014rvg}.
Another useful spectral function we will use in this paper is the zeta function $\zeta_{\rH}(s)$ defined by
\begin{equation}\label{eq:zeta-fun}
    \zeta_{\rH}(s)={\rm Tr}{\rm H}^{-s}=\sum_{n=0}^\infty E_n^{-s},
\end{equation}
which is convergent as ${\rm Re}(s)>C$ for a potential-dependent constant $C$.  
The zeta function and the partition function are related by the Mellin transform
\begin{equation}\label{eq:zeta-int}
    \Gamma(s)\zeta_{\rH}(s)
    =
    \int_0^\infty
    \beta^{s-1}Z(\beta)\,\rd\beta
\end{equation}

In this paper, we focus on the homogeneous potentials of the form $V(x)=x^{N}, -(\ri x)^N$. The first family is the Hermitian homogeneous oscillator, while the second one is the PT-symmetric homogeneous oscillator. Our goal in this paper is to compute the spectral function directly from the ODE/IM correspondence.
In the remainder of this section, we present a brief review of the ODE/IM correspondence to fix our conventions. More details can be found in \cite{Dorey:2007zx}.

\subsection{Counting function and DdV equation}
Let us now consider the Schr{\"o}dinger equation for potentials of generalized anharmonic oscillators
\begin{equation}
    \left(-\frac{\rd^2}{\rd x^2}+x^{2M}+\frac{l(l+1)}{x^2}-E\right)\psi(x)=0,
\end{equation}
where a centrifugal term proportional to the inverse square of the coordinate is introduced. The potential will reduce to homogeneous anharmonic oscillators as long as $l=0$ or $l=-1$. We will assume $M>1$.  Let $y(x,E,l)$ and $\psi(x,E,l)$ be the unique solutions which decay at $x\to \infty$ and $x\to 0$, respectively, with the asymptotics
\begin{equation}\label{eq:sol-ODE}
    \begin{cases}
y(x,E,l)\sim\frac{1}{\sqrt{2\ri}}x^{-M/2}e^{-\frac{x^{M+1}}{M+1}} & x\to\infty\\
\psi(x,E,l)\to x^{l+1} & x\to0
\end{cases}.
\end{equation}
The $T$- and $Q$-functions of the quantum integrable model are defined by the Wronskians
\begin{equation}
  Q(E,l)=W[y(x,E,l),\psi(x,E,l)],\quad  T(E,l)=W[y(\omega x,\omega^{-2} E,l),y(\omega^{-1}x,\omega^2E,l)], 
\end{equation}
where $\omega=\re^{\frac{2\pi\ri}{2M+2}}$. These two functions satisfy the Baxter TQ relation
\begin{equation}\label{eq:TQ}
    T(E)Q(E)=\omega^{l+\frac{1}{2}}Q(\omega^2 E)+\omega^{-l-\frac{1}{2}}Q(\omega^{-2} E).
\end{equation}
At the zeros $E_n$ of $Q(E)$, i.e. $Q(E_n)=0$, one finds the Bethe ansatz equation
\begin{equation}
    1+a(E_n,l)=0,
\end{equation}
where $a(E,l)$ is the counting function defined by
\begin{equation}
a(E,l)=\omega^{2 l+1} \frac{Q\left(\omega^2 E, l\right)}{Q\left(\omega^{-2} E, l\right)}=\omega^{2 l+1} \prod_{n=0}^{\infty}\left(\frac{E_n-\omega^2 E}{E_n-\omega^{-2} E}\right).
\end{equation}

Together with the large $E$ asymptotic behavior of $a(E,l)$, one can convert the Bethe ansatz equation into the Destri–de Vega (DdV) equation \cite{Dorey:2007zx}
\begin{equation}\label{eq:DdV}
    \begin{aligned}
        \ln a(\theta)=&\ri\pi\left(l+\frac{1}{2}\right)-\ri mLe^{\theta}+\int_{-\infty-\ri\delta }^{\infty-\ri\delta }d\theta^{\prime}\varphi\left(\theta-\theta^{\prime}\right)\ln\left(1+a\left(\theta^{\prime}\right)\right)\\
        &-\int_{-\infty+\ri\delta }^{\infty+\ri\delta }d\theta^{\prime}\varphi\left(\theta-\theta^{\prime}\right)\ln\left(1+a\left(\theta^{\prime}\right)^{-1}\right),
    \end{aligned}
\end{equation}
where $E=\re^{\theta/\mu}$ with $\mu=\frac{M+1}{2M}$ and $\delta $ is small positive parameter. With a slight abuse of notation, we have replaced $a(E(\theta))$ by $a(\theta)$ and omitted the $l$-dependence without confusion.
$m L=\frac{\pi^{1/2}}{2M}\frac{\Gamma\left(\frac{1}{2M}\right)}{\Gamma\left(\frac{3}{2}+\frac{1}{2M}\right)}$ is a constant from the zero modes and can be fixed by the large-$\theta$ asymptotic expansion of $\ln a(\theta)$ and the integral kernel is
\begin{equation}\label{eq:kernel}
\varphi(\theta)=\int \frac{d k}{2 \pi} \re^{\ri k \theta} \frac{\sinh \frac{\pi k(1-M)}{2 M}}{2 \sinh \frac{\pi k}{2 M} \cosh \frac{\pi k}{2}}.
\end{equation}

We are especially interested in $l=0$ and $l=-1$ cases, where the centrifugal term vanishes. One can solve the above equations numerically to find the Bethe roots $\theta_n$, satisfying the Bethe ansatz equation $1+a(\theta_n)=0$. These roots correspond to the energy spectrum by $E_n=\re^{\frac{\theta_n}{\mu}}$.



In the following sections, we first express the quantization condition by using the counting function and then present a contour integral formulation for the spectral functions of the Hermitian and PT-symmetric homogeneous oscillators.

\section{Hermitian homogeneous anharmonic oscillator}\label{sec:herm}

In this section, we consider the Hermitian homogeneous oscillator 
\begin{equation}
    \left(-\frac{\rd^2}{\rd x^2}+x^{2M}-E\right)\psi(x)=0,\quad M\in \mathbb{Z}_{>1}
\end{equation}
with the boundary condition: $\psi(x)\to 0$ at $x\to \pm\infty$. Here $l$ is chosen to be $0$ or $-1$.
Since the potential is invariant under the parity: $x\to -x$, the wavefunction decomposes to even/odd sectors $\psi(-x)=\pm \psi(x)$. The boundary condition reduces to
\begin{equation}
    \begin{cases}
\mbox{even sector:} & \psi^{\prime}(0)=0,\quad\psi(\infty)=0\\
\mbox{odd sector:} & \psi(0)=0,\quad\psi(\infty)=0
\end{cases}.
\end{equation}
Recall that $y(x,E,l)$ in \eqref{eq:sol-ODE} goes to zero at infinity, and the $\psi(x,E,l)\sim x^{l+1}$ goes to zero at the origin for $l=0$, which corresponds to the odd sector. For $l=-1$, one instead has  $\psi^\prime (x,E,l)\sim (l+1)x^l$, which goes to zero at the origin, corresponding to the even sector. The quantization condition for the odd/even sector can thus be written by using the Wronskian
\begin{equation}
    \begin{cases}
\mbox{even sector:} & Q(E,-1)=W[y(x,E,-1),\psi(x,E,-1)]=0\\
\mbox{odd sector:} & Q(E,0)=W[y(x,E,0),\psi(x,E,0)]=0
\end{cases},
\end{equation}
which are equivalent to the condition $a(E,-1)+1=0$ and $a(E,0)+1=0$, respectively. To describe the full spectrum, including both odd and even sectors, we define
\begin{equation}
\label{eq:aa-relation}
    \ln(1+{\bf a}(E))=\ln(1+a(E,0))+\ln(1+a(E,-1)),
\end{equation}
whose derivative with respect to $E$ has simple poles at $E_n$. We can thus express the partition function using the contour integral
 \begin{equation}\label{eq:DdV-der}
    \begin{aligned}
         Z(\beta)&=\frac{1}{2\pi\ri}\int_{{\mathcal{C}}}e^{-\beta E}\partial_{E}\ln\left(1+{\bf a}(E)\right)\rd E,
    \end{aligned}
\end{equation}
where the integration contour encircles the positive $E$-axis anticlockwise. Using the $\theta$ variable, one can rewrite the partition function by
\begin{equation}
    \begin{aligned}
       Z(\beta)&=\frac{1}{2\pi\ri}\int_{-\infty-\ri\delta }^{\infty-\ri\delta }\exp\left(-\beta \re^{\frac{\theta}{\mu}}\right)\partial_{\theta}\ln\left(1+{\bf a}(\theta)\right)\rd\theta-\frac{1}{2\pi\ri}\int_{-\infty+\ri\delta }^{\infty+\ri\delta }\exp\left(-\beta \re^{\frac{\theta}{\mu}}\right)\partial_{\theta}\ln\left(1+{\bf a}(\theta)\right)\rd\theta\\&=\frac{1}{2\pi\ri}\int_{-\infty}^{\infty}\exp\left(-\beta \re^{\frac{\theta-\ri\delta }{\mu}}\right)\partial_{\theta}\ln\left(1+{\bf a}(\theta-\ri\delta )\right)\rd\theta-\frac{1}{2\pi\ri}\int_{-\infty}^{\infty}\exp\left(-\beta \re^{\frac{\theta+\ri\delta }{\mu}}\right)\partial_{\theta}\ln\left(1+{\bf a}(\theta+\ri\delta )\right)\rd\theta,
    \end{aligned}
\end{equation}
where we have shifted the variable $\theta$. Integrating the partition function by parts, we obtain
 \begin{equation}
 \label{eq:partition-by-parts}
    \begin{aligned}
       Z(\beta)&=\frac{\beta}{2\pi\ri}\int_{\mathcal{C}}\re^{-\beta E}\ln\left(1+{\bf a}(E)\right)\rd E\\
       &=\frac{\beta}{2\pi\ri\mu}\int_{-\infty}^{\infty}\exp\left(\frac{\theta}{\mu}-\beta \re^{\frac{\theta}{\mu}}\right)\Big(\ln\big(1+{\bf a}(\theta-\ri\delta )\big)-\ln\big(1+{\bf a}(\theta+\ri\delta )\big)\Big)\rd\theta,
    \end{aligned}
\end{equation}
where we have used the shorthand $a(\theta)=a(E(\theta))$ with a slight abuse of notation. From the viewpoint of exact WKB, the two logarithm terms can be regarded as the functions of two lateral Borel resummed WKB periods. Their difference encodes the Stokes discontinuity of the exact quantization condition, which extracts the physical spectrum and yields an unambiguous partition function.

Since $E_n$ are real and positive, it is easy to find ${a}(E,l)^*=\frac{1}{{a}(E^*,l)}$. The partition function thus can be written as
\begin{equation}\label{eq:partition-rewritten}
    \begin{aligned}
        Z(\beta)=&\frac{\beta}{2\pi\ri}\int_{0}^{\infty}\re^{-\beta E}\Big(2\ri{\rm Im}\ln\big[1+{\bf a}(E\re^{\pm\ri\delta/\mu})\big]\pm\ln{\bf a}(E\re^{\mp\ri\delta/\mu})\rd E.\\
        =&\frac{\beta}{2\pi\ri\mu}\int_{-\infty}^{\infty}\exp\left(\frac{\theta}{\mu}-\beta \re^{\frac{\theta}{\mu}}\right)\Big(2\ri{\rm Im}\Big[\ln\big(1+{\bf a}(\theta\pm\ri\delta)\Big]\pm\ln{\bf a}(\theta\mp\ri\delta)\Big)\rd\theta.
    \end{aligned}
\end{equation}

In a similar way, the zeta function $\zeta_{\rH}(s)$ with ${\rm Re}(s)>\mu$  can be computed via the counting function from the DdV equation.
\begin{align}
\zeta_{\rH}(s)
&=
\frac{1}{2\pi \ri}
\int_{\mathcal C}
E^{-s}\,
\partial_E\ln\!\bigl(1+{\bf a}(E)\bigr)\,\rd E
\nonumber\\
&=
\frac{s}{2\pi \ri}
\int_{\mathcal C}
E^{-s-1}\,
\ln\!\bigl(1+{\bf a}(E)\bigr)\,\rd E,
\end{align}
which leads to
\begin{equation}\label{eq:anha-zeta}
    \begin{aligned}
        \zeta_{\rH}(s)=&\frac{1}{2\pi\ri}\int_{-\infty}^{\infty}e^{-s\frac{\theta}{\mu}}\partial_{\theta}\ln\left(1+{\bf a}(\theta-\ri\delta)\right)\rd\theta-\frac{1}{2\pi \ri}\int_{-\infty}^{\infty}e^{-s\frac{\theta}{\mu}}\partial_{\theta}\ln\left(1+{\bf a}(\theta+\ri\delta)\right)\rd\theta\\
       =&\frac{s}{2\pi\ri\mu}\int_{-\infty}^{\infty}e^{-s\frac{\theta}{\mu}}\Big(\ln\left(1+{\bf a}(\theta-\ri\delta)\right)-\ln\left(1+{\bf a}(\theta+\ri\delta)\right)\Big)\rd\theta.
    \end{aligned}
\end{equation}
One can then compute the partition function and spectral $\zeta$-function by plugging the solution of the DdV equation into the above expressions.

\subsection{High-temperature expansions from the DdV equation}

We study the high-temperature limit of the thermal partition function, namely, its small-$\beta$ expansion, using the solution of the DdV equation. This limit is related to the large $E$ behavior of ${a}(E,l)$, which takes the form
\begin{equation}\label{eq:a-expansion}
    \ri\ln a(E,l)
    \sim
    mL E^\mu
    -\pi\left(l+\frac{1}{2}\right)
    +I_{1,l}E^{-\mu}
    +I_{3,l}E^{-3\mu}
    +\cdots.
\end{equation}
Or equivalently, in terms of the rapidity variable $\theta$,
\begin{equation}
    \ri\ln a(\theta,l)
    \sim
    mL\re^\theta
    -\pi\left(l+\frac{1}{2}\right)
    +I_{1,l}\re^{-\theta}
    +I_{3,l}\re^{-3\theta}
    +\cdots.
\end{equation}
The odd-exponential structure of the
above expansion holds for integer $M$. As for the kernel \eqref{eq:kernel}, the poles of the factor $1/\cosh(\pi k/2)$ occur at
$k=\ri(2n+1)$ and generate the terms
$\re^{-(2n+1)\theta}$.\footnote{For noninteger $M$, the factor $1/\sinh\left(\frac{\pi k}{2M}\right)$
can generate an additional family of poles at $k=2\ri Mn$, associated
with dual nonlocal integrals of motion. For integer $M$, however, these
poles are cancelled by zeros of the numerator.} 
Therefore, for the integer homogeneous oscillators considered here, the
large-$\theta$ expansion contains only the odd exponentials
$\re^{-(2n+1)\theta}$. Although some of their coefficients may vanish at special
integer values of $M$, this will be clarified later.

We consider the high-temperature expansion of the partition function from \eqref{eq:partition-by-parts} in the $E$ variable. In this limit, the term $2\ri {\rm Im} \ln\left(1+{a}(E\re^{-\ri\delta/\mu},l)\right)$ is exponentially suppressed at large $E$. So we can capture the small $\beta$ asymptotic expansion by accounting for
\begin{equation}
    Z(\beta)
    \sim\frac{\beta}{2\pi}
    \int_0^\infty
    \re^{-\beta E}\ri \ln{\bf a}(\theta)\rd E=
    \frac{\beta}{2\pi}
    \int_0^\infty
    \re^{-\beta E}(\ri \ln {a}(E,0)+\ri\ln a(E,-1))\rd E.
\end{equation}
Inserting the large-$E$ expansion of $\ln {a}(E,l)$ \eqref{eq:a-expansion} and using \footnote{We note that the integral \eqref{eq:small-beta-integral} is convergent at
$E=0$ when $s_n<1$. When $s_n>1$, the equality should be understood as analytic continuation.}
\begin{equation}
\label{eq:small-beta-integral}
    \frac{\beta}{2\pi}
    \int_0^\infty
    \re^{-\beta E}E^{-s_n}\,\rd E
    \overset{\rm{ac}}{=}
    \frac{
        \Gamma\left(1-s_n\right)
    }{2\pi}
    \beta^{s_n},\qquad s_n=(2n+1)\mu.
\end{equation}
we obtain the small-$\beta$ expansion as follows:
\begin{equation}
\begin{aligned}
    Z(\beta)=
    A_{-1}\beta^{-\mu}
    +A_1\beta^\mu
    +A_3\beta^{3\mu}
    +\cdots.
\end{aligned}
\end{equation}
Defining
\begin{equation}
    I_{2n+1}
    \equiv
    I_{2n+1,0}+I_{2n+1,-1},
\end{equation}
the coefficients are given by
\begin{equation}
    \label{eq:ai-relation}
    A_{2n+1}
    =
    \frac{
        I_{2n+1}
        \Gamma\left(1-s_n\right)
    }{2\pi},
    \qquad n=-1,0,1,\cdots,
\end{equation}
where $I_{-1}=I_{-1,0}+I_{-1,-1}=2mL$.
On the other hand, the high-temperature expansion can be computed
independently using the Wigner--Kirkwood expansion. For example,
\begin{equation}\label{eq:a-coefficients}
\begin{aligned}
     A_{-1}&=\frac{1}{\sqrt{\pi}}\Gamma\left(1+\frac{1}{2M}\right),\\
     A_1&=-\frac{M}{6\sqrt{\pi}}
    \Gamma\left(2-\frac{1}{2M}\right),\\
    A_3&=\frac{\left(8 M^3+2 M^2-3 M\right) \Gamma \left(2-\frac{3}{2 M}\right)}{720 \sqrt{\pi }}\\
    A_5&=\frac{\left(-384 M^5+16 M^4+556 M^3+44 M^2-139 M\right) \Gamma \left(2-\frac{5}{2
   M}\right)}{181440 \sqrt{\pi }}.
\end{aligned}
\end{equation}
We can therefore determine the small-$\beta$ expansion of the
partition function from the large-$\theta$ asymptotic expansion of the DdV solution, and confirm the validity of relation \eqref{eq:ai-relation} against the Wigner--Kirkwood expansion in \cite{Jizba:2014rvg}.

Moreover, the Mellin representation \eqref{eq:zeta-int}
implies that $\Gamma(s)\zeta_{\rH}(s)$ has poles at $s=-s_n$. When $s_n \notin \mathbb{Z}_{\geq 0}$, $\Gamma(s)$ is finite at that point, these are
also poles of $\zeta_{\rH}(s)$, with residues
\begin{equation}
\begin{aligned}
    \operatorname*{Res}_{s=-s_n}\zeta_{\rH}(s)=
    \frac{A_{2n+1}}{\Gamma(-s_n)}.
\end{aligned}
\end{equation}
When $s_n\in\mathbb{Z}_{\geq 0}$, the Gamma factor in \eqref{eq:ai-relation} is divergent, meanwhile $I_{2n+1}$ vanishes, and the relation  \eqref{eq:ai-relation} holds in $s\to s_n$ limit. This implies that
\begin{equation}
    \zeta_{\rH}(-s_n)=(-1)^{s_n} s_n!A_{2n+1},
\end{equation}
which must be understood in the sense of analytic continuation of $\zeta_{H}(s)$. For example, in the sextic potential with $M=3$, when $n=1$, $s_n=2$, we conclude
\begin{equation}
    \zeta_{\rH}(-2)=2A_3=\frac{5}{16}.
\end{equation}

\subsection{Quartic potential}
In this subsection, we illustrate our general analysis with the quartic potential with $M=2$, whose kernel \eqref{eq:kernel} simplifies to
\begin{equation}
    \varphi(\theta)=-\frac{1}{2\pi\cosh\theta}
\end{equation}
We solve these DdV equations using the discretized Fourier transformation, and obtain the solutions ${a}(\theta\pm\ri\delta,l)$ for $l=-1$ and $l=0$. Substituting these into \eqref{eq:partition-by-parts}, we obtain the partition function $Z(\beta)$.

In Table \ref{tab:E0-com}, we compare the ground state energy via the partition function \eqref{eq:E0-par} and the one obtained by the Hamiltonian diagonalization, which shows agreement to high precision.
\begin{table}[htp]
    \centering
    \begin{tabular}{ccc}
     \toprule
        Method & Partition function& Diagonalization \\\hline
       $E_0$  & $1.060362090484065$  & $1.060362090484183$\\
        \toprule
    \end{tabular}
    \caption{Ground-state energy extracted from Eq. \eqref{eq:E0-par} at $\beta=10$ and obtained by Hamiltonian diagonalization.}
    \label{tab:E0-com}
\end{table}
We also compare the partition function for finite $\beta$ with those from the Hamiltonian diagonalization method to find great agreement.

\begin{figure}
    \centering
    \includegraphics[width=0.7\linewidth]{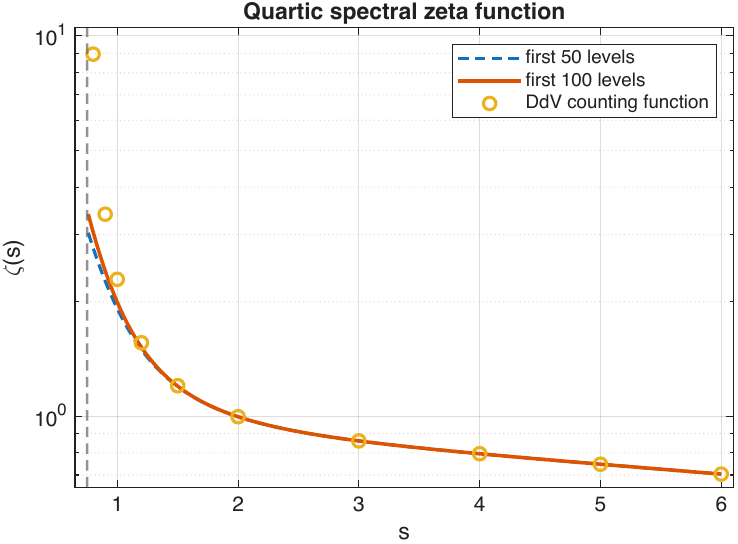}
    \caption{Quartic spectral zeta function from the DdV counting function compared with partial sums over the first \(50\) and \(100\) energy levels. The vertical dashed line marks \(s=3/4\), and the vertical axis is logarithmic.}
    \label{fig:zeta-plot}
\end{figure}

To calculate the $\zeta_{\rH}$ function, we decompose the integral into three parts $\int_{-\infty}^\infty=\int_{-\infty}^{\theta_{\rm min}}+\int_{\theta_{\rm min}}^{\theta_{\rm max}}+\int^{\infty}_{\theta_{\rm max}}$. The second term can be computed using the solution of the DdV equation. The first one will always vanish by choosing small enough $\theta_{\rm min}$. The third one can be approximated by using the large $E$ expansion (WKB expansion). In Table \ref{tab:zeta-com-v1}, we compare our zeta function with the results in \cite{Voros1983}. We also show the comparison of the zeta function from our approach with the partial sum over the first several levels in Figure \ref{fig:zeta-plot}.


\begin{table}[t]
\centering
\begin{tabular}{ccc}
\toprule
$s$ &
$\zeta^{\mathrm{DdV}}(s)$ &
$\zeta^{\mathrm{Voros}}(s)$ \\
\midrule
1 & 2.289908804320 & 2.289908804320 \\
2 & 0.996320827678 & 0.996320827679 \\
3 & 0.860517138943 & 0.860517138943 \\
4 & 0.796211192708 & 0.796211192704 \\
5 & 0.747295110986 & 0.747295110967 \\
6 & 0.703855987778 & 0.703855987715 \\
\bottomrule
\end{tabular}
\caption{
Spectral zeta function for the quartic oscillator obtained directly
from the DdV counting functions by integration by parts. We integrate in the interval $[-30,30)$ by $N_\theta=2^{15}$ points and take $\delta=2^{-6}$. The leading contribution beyond the upper rapidity cutoff is included by the WKB approximation. The last column lists the corresponding values in \cite{Voros1983}.
}
\label{tab:zeta-com-v1}
\end{table}

For the quartic oscillator, the expansion of the partition function takes the form of
\begin{equation}
\begin{aligned}
    Z(\beta)\sim{}
    \frac{\Gamma(1/4)}{4\sqrt{\pi}}\,
    \beta^{-3/4}
    -\frac{\Gamma(3/4)}{4\sqrt{\pi}}\,
    \beta^{3/4}
    +\frac{11\Gamma(1/4)}{480\sqrt{\pi}}\,
    \beta^{9/4}-\frac{61\Gamma(3/4)}{1440\sqrt{\pi}}\,
    \beta^{15/4}
    +O\left(\beta^{21/4}\right).
\end{aligned}
\end{equation}
Comparing these coefficients with those extracted from the DdV
integrals of motion, we find excellent agreement, with the largest relative error being of order $10^{-13}$. Moreover, we obtain
\begin{equation}
\begin{gathered}
    \operatorname*{Res}_{s=3/4}\zeta_{\rH}(s)
    =
    \frac{
        \Gamma(1/4)
    }{
        4\sqrt{\pi}\,\Gamma(3/4)
    },
    \\
    \operatorname*{Res}_{s=-3/4}\zeta_{\rH}(s)
    =
    \frac{
        3\Gamma(3/4)
    }{
        16\sqrt{\pi}\,\Gamma(1/4)
    }.
\end{gathered}
\end{equation}
These values are consistent with those of
\cite{Voros:1979xj} after taking into account the normalization of the
Hamiltonian.

\section{PT-symmetric anharmonic oscillator}\label{sec:PT}
We study the PT-symmetric anharmonic oscillator
\begin{equation}\label{eq:PT-anha}
 \Big(-\frac{\dd^2}{\dd q^2}-(\ri q)^{2M}-E^{\rm PT}\Big)\Psi(q)=0,\quad 2M\in \mathbb{Z}_{>1}.
\end{equation}
Generically, the coordinate $q$ is not restricted to the real axis. The wavefunction should be defined along a PT-symmetric contour ${\cal C}$ in the complex $q$-plane 
\begin{equation}
    \Psi(q)\to 0\quad \mbox{as} \quad q\to \infty\, \mbox{along the two ends of}\,\, {\cal C},
\end{equation}
Here the two ends of the contour ${\cal C}$ should lie inside a pair of Stokes sectors. Each Stokes sector has angular width $\frac{2\pi }{2M+2}$. The central directions of the two sectors are \cite{Bender:1998ke}
\begin{equation}
   {\rm Arg}(q)= -\frac{\pi}{2}\pm\frac{\pi}{M+1}
\end{equation}
respectively. These two sectors can also be written as
\begin{equation}
    \begin{aligned}
        &\mathcal{S}_{R}^{\rm PT}:\qquad\left|\arg(q)+\frac{\pi}{2}-\frac{\pi}{M+1}\right|<\frac{\pi}{2(M+1)}\\&\mathcal{S}_{L}^{\rm PT}:\qquad\left|\arg(q)+\frac{\pi}{2}+\frac{\pi}{M+1}\right|<\frac{\pi}{2(M+1)}.
    \end{aligned}
\end{equation}
The boundary condition can thus be expressed as
\begin{equation}
    \Psi(q)\to 0\quad \mbox{as}\quad q\to \infty \,\,\mbox{inside}\, {\cal S}_L^{\rm PT}
\,\mbox{and} \,{\cal S}_R^{\rm PT}.
\end{equation}

To take advantage of the ODE/IM correspondence, we let $x=\ri q$. Equation \eqref{eq:PT-anha} thus becomes
\begin{equation}
    \Big(-\frac{\dd^{2}}{\dd x^{2}}+x^{2M}-E\Big)\psi(x)=0,
\end{equation}
where $E=-E^{\rm PT}$. In the $x$ variable, the boundary condition becomes
\begin{equation}
    \psi(x)\to0\quad\mbox{as}\quad x\to\infty\,\mbox{inside}\,{\cal S}_{L}\,\mbox{and }\,{\cal S}_{R},
\end{equation}
where the two sectors are given by
\begin{equation}
    \mathcal{S}_{R,L}:\quad\left|\arg(x)\mp\frac{\pi}{M+1}\right|<\frac{\pi}{2(M+1)}.
\end{equation}
In the sector $\mathcal{S}_{R,L}$, the unique decaying solutions are $y(\omega x,\omega^{-2} E,l)$ and $y(\omega^{-1}x,\omega^2E,l)$, respectively. The quantization condition thus becomes
\begin{equation}
    T(E,0)=W[y(\omega x,\omega^{-2} E,0),y(\omega^{-1}x,\omega^2E,0)]=0,
\end{equation}
where $l=-1$ is supposed to give the same spectrum. In the PT-variable, the quantization becomes \cite{Dorey:2004fk}
\begin{equation}
    T(-E^{\rm PT},0)=0,
\end{equation}
where the eigenvalues $E_n^{\rm PT}$ are real and positive \cite{Bender:1998ke,Dorey:2001uw}. From the Baxter TQ relation, this condition can be rewritten as
\begin{equation}
    1+a(-E_{n}^{\rm PT})=0.
\end{equation}
The partition function thus can be expressed by
\begin{equation}
    Z^{\rm PT}(\beta)=\sum_{n=0}^\infty \re^{-\beta E_n^{\rm PT}}=\frac{1}{2\pi\ri}\int_{{\mathcal{C}}}e^{-\beta E^{{\rm PT}}}\partial_{E^{{\rm PT}}}\ln\left(1+a(-E^{{\rm PT}})\right)dE^{{\rm PT}}.
\end{equation}
Introducing the rapidity-like variable by $E^{\rm PT}=e^{\theta^{\rm PT}/\mu}$, the quantization becomes
\begin{equation}\label{eq:PT-EQC-theta}
    1+a(\theta^{{\rm PT}}+\mu\pi\ri)=0
\end{equation}
with a slight abuse of notation.
Note that the T-/Q-function and the counting function are defined for the $x$-variable. Since $E^{\rm PT}$ is real and positive, we have to shift $\theta$ from the real axis $\theta^{\rm PT}$ to the line ${\rm Im}(\theta)=\mu\pi$ in the quantization condition, which goes beyond the strip for \eqref{eq:DdV}. We should use the second determination by picking up the contribution of the pole at $\theta=\frac{\pi\ri}{M}$ in the kernel $\varphi(\theta)$ \cite{Dorey:2004fk}
\begin{equation}\label{eq:DdV-2nd}
    \begin{aligned}
        \ln a(\theta^{{\rm PT}}+\mu\pi\ri)=&2\ri\sin(\frac{\pi}{2M})mLe^{\theta^{{\rm PT}}}+\int_{-\infty-\ri\delta }^{\infty-\ri\delta }\rd\theta^{\prime}\varphi_{{\rm II}}\left(\theta^{{\rm PT}}+\mu\pi\ri-\theta^{\prime}\right)\ln\left(1+a\left(\theta^{\prime}\right)\right)\\&-\int_{-\infty+\ri\delta }^{\infty+\ri\delta }\rd\theta^{\prime}\varphi_{{\rm II}}\left(\theta^{{\rm PT}}+\mu\pi\ri-\theta^{\prime}\right)\ln\left(1+a\left(\theta^{\prime}\right)^{-1}\right),
    \end{aligned}
\end{equation}
where $\varphi_{{\rm II}}(\theta)=\varphi(\theta)-\varphi(\theta-\frac{\ri\pi}{M})$. 

We thus should solve \eqref{eq:DdV} at first, and plug the resulting solution into the \eqref{eq:DdV-2nd} to obtain $a(\theta^{\rm PT}+\mu\pi\ri)$. Substituting this into the partition function, we obtain the integral representation of the partition function for the PT-symmetric anharmonic oscillator
\begin{equation}\label{eq:PT-par}
   \begin{aligned}
     Z^{\rm PT}(\beta)=&\frac{\beta}{2\pi\ri\mu}\int_{-\infty}^{\infty}\exp\left(\theta^{{\rm PT}}/\mu-\beta\re^{\theta^{{\rm PT}}/\mu}\right)\Big(\ln\left(1+a(\theta^{{\rm PT}}+\mu\pi\ri-\ri\delta)\right)-\ln\left(1+a(\theta^{{\rm PT}}+\mu\pi\ri+\ri\delta)\right)\Big)\rd\theta^{{\rm PT}},
     \end{aligned}
\end{equation}
where we have used integration by parts. The zeta function can be computed similarly using the integral by
\begin{equation}\label{eq:PT-zeta}
    \begin{aligned}
        \zeta_{\rH}^{\rm PT}(s)=&\frac{s}{2\pi\ri\mu}\int_{-\infty}^{\infty}e^{-s\theta^{{\rm PT}}/\mu}\Big(\ln\left(1+a(\theta^{{\rm PT}}+\mu\pi\ri-\ri\delta)\right)-\ln\left(1+a(\theta^{{\rm PT}}+\mu\pi\ri+\ri\delta)\right)\Big)\rd\theta^{{\rm PT}}.
    \end{aligned}
\end{equation}

\subsection{High-temperature expansion}

We now consider the high-temperature expansion of the PT-symmetric partition function. Similar to the Hermitian case, we still restrict to integer
$M$ \footnote{For odd integer $M$, the PT-oscillator $p^2-(\ri q)^{2M}$ reduces algebraically to \(p^2+q^{2M}\). If the boundary conditions are imposed on the real axis, the resulting spectral problem coincides with the Hermitian oscillator. However, our boundary conditions for PT-oscillator generally select a different pair of asymptotic sectors and hence define a distinct spectral problem, except for $M=1$.}, for which the dual nonlocal pole family of the DdV kernel is
cancelled. The large-$E$ expansion of the counting function introduced
above therefore takes the form
\begin{equation}
\label{eq:pt-large-E-expansion}
\ri \ln a\left(-E^{\mathrm{PT}}\right) \sim B_M^{\mathrm{PT}}\left(E^{\mathrm{PT}}\right)^\mu+J_1\left(E^{\mathrm{PT}}\right)^{-\mu}+J_3\left(E^{\mathrm{PT}}\right)^{-3 \mu}+\cdots,
\end{equation}
where $E^{\rm PT}=-E$, and 
\begin{equation}
\label{eq:pt-leading-action}
    B_M^{\mathrm{PT}}
    =
    2mL\sin\left(\frac{\pi}{2M}\right).
\end{equation}
The factor $\sin(\pi/(2M))$ arises from the complex Stokes contour. Using the same contour manipulation as in the Hermitian case, we find
\begin{equation}
\label{eq:pt-smooth-partition}
    Z^{\mathrm{PT}}(\beta)
    \sim
    \frac{\beta}{2\pi}
    \int_0^\infty
    \re^{-\beta E^{\rm PT}}
    \ri\ln a(-E^{\rm PT})\,\rd E^{\rm PT}.
\end{equation}
Unlike the Hermitian full-line problem, the PT-symmetric case contains a single sector $l=0$. Substituting \eqref{eq:pt-large-E-expansion} into
\eqref{eq:pt-smooth-partition}, we obtain
\begin{equation}
\label{eq:pt-small-beta-expansion}
    Z^{\mathrm{PT}}(\beta)
    \sim
    A_{-1}^{\mathrm{PT}}\beta^{-\mu}
    +A_1^{\mathrm{PT}}\beta^\mu
    +A_3^{\mathrm{PT}}\beta^{3\mu}
    +A_5^{\mathrm{PT}}\beta^{5\mu}
    +\cdots.
\end{equation}
Defining
\begin{equation}
    s_n=(2n+1)\mu,
    \qquad
    J_{-1}\equiv B_M^{\mathrm{PT}},
\end{equation}
the coefficients are related by
\begin{equation}
\label{eq:pt-ai-relation}
    A_{2n+1}^{\mathrm{PT}}
    =
    \frac{
        J_{2n+1}\Gamma\left(1-s_n\right)
    }{2\pi},
    \qquad
    n=-1,0,1,\cdots.
\end{equation}
In particular,
\begin{equation}
\label{eq:pt-leading-heat-coefficient}
\begin{aligned}
    A_{-1}^{\mathrm{PT}}
    &=
    \frac{
        B_M^{\mathrm{PT}}\Gamma(1+\mu)
    }{2\pi}=
    \sin\left(\frac{\pi}{2M}\right)
    \frac{1}{\sqrt{\pi}}
    \Gamma\left(1+\frac{1}{2M}\right).
\end{aligned}
\end{equation}

The coefficients can be checked independently using the
Wigner--Kirkwood expansion continued to the
PT-symmetric contour. If $A_{2n+1}$ denotes the corresponding
Hermitian coefficient, the contour rotation gives
\begin{equation}
\label{eq:pt-wk-relation}
    A_{2n+1}^{\mathrm{PT}}
    =
    \sin\left(\frac{\pi (2n+1)}{2}\right)
    \sin\left(\frac{\pi (2n+1)}{2M}\right)
    A_{2n+1},
    \qquad
    n=-1,0,1,\cdots.
\end{equation}
Equations \eqref{eq:pt-ai-relation} and
\eqref{eq:pt-wk-relation} provide a direct comparison between the
large-$\theta$ coefficients of the DdV solution and the
high-temperature coefficients of the PT-symmetric
partition function. At special integer values for which
$s_n\in\mathbb{Z}_{\geq0}$, these relations are understood by taking the limit, as in the Hermitian case. The Mellin representation gives
\begin{equation}
    \operatorname*{Res}_{s=\mu}
    \zeta^{\rm PT}_{\mathrm{H}}(s)
    =
    \frac{
        A_{-1}^{\mathrm{PT}}
    }{
        \Gamma(\mu)
    },
\end{equation}
and, for $s_n\notin\mathbb{Z}_{\geq0}$,
\begin{equation}
    \operatorname*{Res}_{s=-s_n}
    \zeta^{\rm PT}_{\mathrm{H}}(s)
    =
    \frac{
        A_{2n+1}^{\mathrm{PT}}
    }{
        \Gamma(-s_n)
    }.
\end{equation}
When $s_n\in\mathbb{Z}_{\geq0}$, the corresponding point is regular
and instead satisfies
\begin{equation}
    \zeta^{\rm PT}_{\mathrm{H}}(-s_n)
    =
    (-1)^{s_n}s_n!\,
    A_{2n+1}^{\mathrm{PT}}.
\end{equation}
 
\subsection{PT-symmetric cubic oscillator}
In this subsection, we study the PT-symmetric cubic oscillator in detail
\begin{equation}
 \Big(-\frac{\dd^2}{\dd q^2}+\ri q^3-E^{\rm PT}\Big)\Psi(q)=0.
  \label{eq:PT-cubic}
\end{equation}
Since $q\to \pm \infty$ lie in the sectors $\mathcal{S}_{R,L}^{\rm PT}$ respectively, the boundary condition reduces to the standard one
\begin{equation}
    \Psi(q)\to 0\quad \mbox{as}\quad q\to \pm\infty.
\end{equation}
We compute the partition function \eqref{eq:PT-par} and zeta function \eqref{eq:PT-zeta} by solving the DdV equation. In Table \ref{tab:PT-E0-com}, we compare the ground state energy $E_0^{\rm PT}$ obtained from the partition function with the one obtained by the Hamiltonian diagonalization. In Table \ref{tab:zeta-com-cubic}, we compare our $\zeta^{\rm PT}_{\rH}$ function with the known results in the literature.

\begin{table}[htp]
    \centering
    \begin{tabular}{ccc}
     \toprule
        Method & Partition function& Diagonalization \\\hline
       $E_0^{\rm PT}$  & $1.156267071987999$  & $1.156267071988073$\\
        \toprule
    \end{tabular}
    \caption{Ground-state energy extracted from Eq. \eqref{eq:E0-par} at $\beta=10$ and obtained by Hamiltonian diagonalization.}
    \label{tab:PT-E0-com}
\end{table}

\begin{table}[htp]
    \centering
    \small
    \begin{tabular}{ccccc}
        \toprule
        $s$
        & $\zeta_{\rm H}^{\rm PT}(s)$
        & $\zeta_{\rm H}^{\rm PT, Known}(s)$
        \\
        \midrule
        $1$
        & $2.835094933971718$
        & $2.835094933971790$\cite{Mezincescu:2000fd}
        \\
        $2$
        & $0.848569248036258$
        & $0.848569248036396$ \cite{Watkins:2011gx}
        \\
        \bottomrule
    \end{tabular}
    \caption{Spectral zeta values obtained directly from Eq. \eqref{eq:PT-zeta} using the DdV solution. We use the same numerical setup as in Table \ref{tab:zeta-com-v1}. The reference values in the last column are taken from \cite{Mezincescu:2000fd,Watkins:2011gx}.
    }
    \label{tab:zeta-com-cubic}
\end{table}

\section{Conclusion and discussion}
In this paper, we have presented an integrable model description for the partition function and the zeta function of the homogeneous Hermitian and PT-symmetric oscillator based on the ODE/IM correspondence. For the Hermitian case, the parity symmetry decomposes the full spectral problem into even and odd sectors. In the PT-symmetric case, we need to choose a pair of PT-related Stokes sectors. In both cases, the quantization condition can be expressed using the counting function $a(E)$, which can be solved via the DdV equation of the integrable model. Using this counting function, we have provided a contour integral representation of the partition function and the zeta function. We have also studied the asymptotic expansion of the partition function and derived several exact results for the spectral zeta function. Further exact values can be obtained using the T-Q relation, as demonstrated in \cite{Watkins:2011gx, Voros:2022yos, Kamata:2025dkk}.

One can also rewrite the quantization function in terms of T-/Y-functions, which satisfy the TBA equations of the integrable model. However, this has to be done case by case. In this work, we have mostly focused on the pure oscillator, i.e., $l=0,-1$. It is straightforward to generalize the procedure to arbitrary $l$. One of the most interesting directions is to generalize to an arbitrary polynomial potential, whose associated Riemann-Hilbert problem of WKB periods can be reformulated as a TBA system. Although the direct computation of the spectral problem and the partition function is more complicated, it encodes richer resurgent structure at the same time. Related developments and attempts can be found in \cite{Masoero:2010is,Fioravanti:2020udo,vanSpaendonck:2023znn,FRS}.


\subsection*{Acknowledgements}
We would like to thank Davide Fioravanti, Jie Gu, Katsushi Ito, Syo Kamata, Yong Li, Marco Rossi and Roberto Tateo for useful discussions. H.S. is supported by the National Natural Science Foundation of China (Grant No.12405087), Henan Postdoc Foundation (Grant No.22120055) and the Startup Funding of Zhengzhou University (Grant No.121-35220049, 121-35220581). J.Y. is supported by the National Natural Science Foundation of China (Grant No.12247103) and JST SPRING, Japan Grant Number JPMJSP2106. J.Y. would like to thank Zhengzhou University for its kind hospitality.

\bibliographystyle{JHEP}






\end{document}